\documentclass[aps,prx,twocolumn,floatfix,superscriptaddress,amsmath,amssymb,longbibliography]{revtex4-1}

\usepackage{graphicx}
\usepackage{dcolumn}%
\usepackage{bm}
\usepackage[pdftex]{color}
\usepackage[dvipsnames]{xcolor}
\usepackage[colorlinks,bookmarks=false,citecolor=RoyalBlue,linkcolor=BrickRed,urlcolor=OliveGreen]{hyperref}
\usepackage[mathlines]{lineno}
\usepackage{braket}
\usepackage{bbm}
\usepackage{verbatim}
\usepackage{ulem}

\newcommand{\Z}{{\ensuremath{\mathbb{Z}}}}

\def\ie{\begin{equation}\begin{aligned}}
\def\fe{\end{aligned}\end{equation}}

\begin{document}

\title{Disentangling critical quantum spin chains with Clifford circuits}

\author{Chaohui Fan}
\affiliation{Beijing National Laboratory for Condensed Matter Physics, Institute of Physics, Chinese Academy of Sciences, Beijing 100190, China}
\affiliation{School of Physical Sciences, University of Chinese Academy of Sciences, Beijing 100049, China}

\author{Xiangjian Qian}
\affiliation{Key Laboratory of Artificial Structures and Quantum Control (Ministry of Education),  School of Physics and Astronomy, Shanghai Jiao Tong University, Shanghai 200240, China} 

\author{Hua-Chen Zhang} \thanks{huachen.zhang@ift.csic.es}
\affiliation{Instituto de F\'isica Te\'orica UAM/CSIC, C/ Nicol\'as Cabrera 13-15, Cantoblanco, 28049 Madrid, Spain}

\author{Rui-Zhen Huang} \thanks{huangrzh@icloud.com}
\affiliation{Department of Physics and Astronomy, University of Ghent, 9000 Ghent, Belgium}
\affiliation{Graduate School of China Academy of Engineering Physics, Beijing 100193, China}

\author{Mingpu Qin} \thanks{qinmingpu@sjtu.edu.cn}
\affiliation{Key Laboratory of Artificial Structures and Quantum Control (Ministry of Education),  School of Physics and Astronomy, Shanghai Jiao Tong University, Shanghai 200240, China}

\affiliation{Hefei National Laboratory, Hefei 230088, China}

\author{Tao Xiang} \thanks{txiang@iphy.ac.cn}
\affiliation{Beijing National Laboratory for Condensed Matter Physics, Institute of Physics, Chinese Academy of Sciences, Beijing 100190, China}
\affiliation{School of Physical Sciences, University of Chinese Academy of Sciences, Beijing 100049, China}
\affiliation{Beijing Academy of Quantum Information Sciences, Beijing 100193, China}

\date{\today}

\begin{abstract}
Clifford circuits can be utilized to disentangle quantum states with polynomial cost, thanks to the Gottesman-Knill theorem. Based on this idea, the Clifford circuits augmented matrix product states (CAMPS) method, which is a seamless integration of Clifford circuits within the density-matrix renormalization group algorithm, was proposed recently and was shown to be able to reduce entanglement in various quantum systems. In this work, we further explore the power of the CAMPS method in critical spin chains described by conformal field theories (CFTs) in the scaling limit. We find that the optimized disentanglers correspond to {\it duality} transformations, which significantly reduce the entanglement entropy in the ground state. For the critical quantum Ising spin chain governed by the Ising CFT with self-duality, the Clifford circuits found by CAMPS coincide with the duality transformation, i.e., the Kramers-Wannier self-duality in the critical Ising chain. It reduces the entanglement entropy by mapping the free conformal boundary condition to the fixed one. In the more general case of the XXZ chain, the CAMPS gives rise to a duality transformation mapping the model to the quantum Ashkin-Teller spin chain. Our results highlight the potential of the framework as a versatile tool for uncovering hidden dualities and simplifying the entanglement structure of critical quantum systems.     
\end{abstract}

\maketitle

\section{introduction}
It is generally believed that the classical simulation of quantum many-body systems or quantum circuits is hard, but Clifford circuits made solely of Clifford gates (Hadamard, S, and Controlled-NOT gates) \cite{Nielsen_Chuang_2010} can be efficiently simulated classically according to the Gottesman-Knill theorem \cite{gottesman1997stabilizer,PhysRevA.70.052328,PhysRevA.73.022334}. Even though Clifford gates are not universal in quantum computing, the state constructed from Clifford gates, i.e., the stabilizer state, can host large entanglement \cite{PhysRevX.7.031016}. In the past few decades, many methods based on tensor network states~\cite{PhysRevLett.69.2863,PhysRevLett.102.180406,PhysRevB.81.165104,SCHOLLWOCK201196,ORUS2014117,RevModPhys.93.045003,xiang2023density} were proposed to simulate quantum many-body systems. Given that the power of tensor networks is bounded by the entanglement entropy the underlying ansatz can support, it is very tempting to try to combine tensor network states [e.g., matrix product states (MPSs)] and Clifford circuits to leverage the advantages of both of them \cite{doi:10.1021/acs.jctc.3c00228, masotllima2024stabilizertensornetworksuniversal,lami2024quantum}. The key is then to find an efficient method to optimize the combined ansatz. 

\begin{figure}[!tbp]
    \centering
    \includegraphics[width=0.9\linewidth]{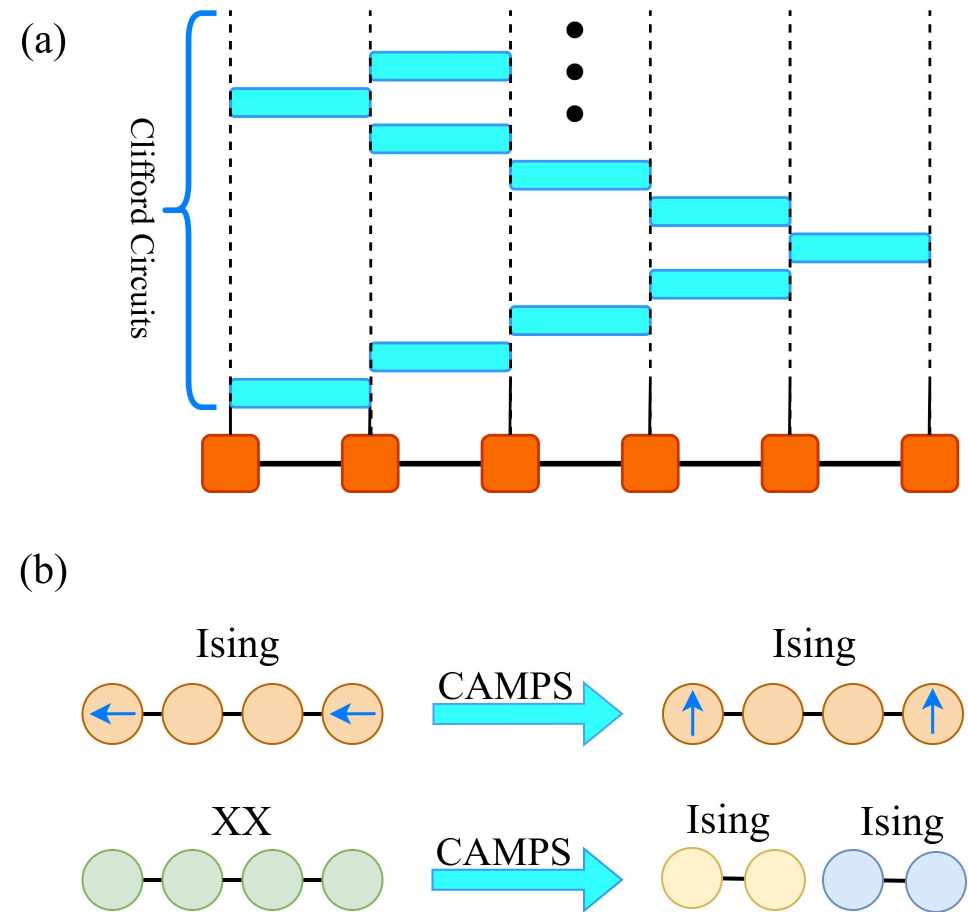}
    \caption{(a) Schematic illustration of the wave-function ansatz in the CAMPS method. In CAMPS, two-site Clifford circuits are applied to the MPS repeatedly. Details about the optimization of the ansatz can be found in Ref.~\cite{PhysRevLett.133.190402}. (b) Critical spin chains conjugated by Clifford gates obtained from variational optimization in CAMPS. CAMPS changes the free boundary condition of the quantum Ising chain to the fixed boundary condition~\cite{ref_ising_bc_discussion}, and maps the XX chain to two decoupled critical quantum Ising chains. The XXZ chain is mapped to the Ashkin-Teller quantum spin chain.}
    \label{fig:1}
\end{figure}

Clifford circuits augmented matrix product states (CAMPS) method \cite{PhysRevLett.133.190402}, which is a seamless integration of Clifford circuits within the density-matrix renormalization group (DMRG) algorithm, was proposed recently and was shown to be very efficient and able to reduce entanglement in various quantum systems \cite{PhysRevLett.133.190402} [an illustration of the wave-function ansatz of CAMPS can be found in Fig.~\ref{fig:1}(a)]. Shortly after \cite{PhysRevLett.133.190402}, the CAMPS method was generalized to the calculation of time evolution \cite{2024arXiv240703202Q,2024arXiv240701692M} and finite temperature \cite{2024arXiv241015709Q} in the framework of time-dependent variational principle \cite{PhysRevLett.107.070601,PhysRevB.94.165116}, demonstrating its power to improve the accuracy significantly with mild overhead in these cases. Accompanying the proposal of the CAMPS method is the concept of non-stabilizerness entanglement entropy (NsEE) \cite{2024arXiv240916895H}, which is essentially the entanglement entropy in a quantum state that cannot be removed with Clifford circuits. NsEE is shown to be a measurement of the hardness for simulating the quantum system classically, better than either entanglement entropy or non-stabilizerness/magic alone. 

In the previous studies, the focus was mostly on the disentangling power of CAMPS, i.e., the ability to reduce the entanglement in the targeted state and to improve the accuracy accordingly. In this work, we carry out an in-depth investigation of the structure of the resulting Clifford circuits and the conjugated Hamiltonian obtained by CAMPS. We focus on one-dimensional critical chains that can be described by conformal field theories (CFTs) in the scaling limit. It is known that for quantum critical chains with open boundary conditions, the bipartite entanglement entropy in ground state scales logarithmically with system size as $S = \frac{c}{6}\ln L + b$ \cite{Cardy2016,PhysRevLett.96.100603} with $c$ the central charge of the underlying CFT and $b$ containing the contribution from boundaries~\cite{PhysRevLett.96.100603}. In this sense, critical chains are more difficult to simulate than gapped ones, which have finite entanglement entropy in the ground state. Moreover, critical chains admit richer structures to explore. The question we want to answer in this work is twofold. On one hand, we would like to see how much entanglement can be removed by Clifford circuits for typical critical spin chains and to investigate the features of the resulting Clifford circuits and the conjugated Hamiltonians; on the other hand, we aim to figure out how the underlying CFT transforms under the application of Clifford circuits using the CAMPS method.

In this work, we study the critical quantum Ising chain and the XXZ chain.
Interestingly, we find that CAMPS yields the exact duality transformations for both models. For the critical quantum Ising chain, CAMPS produces the Kramers-Wannier duality transformation and changes the free boundary condition to the fixed one [see Fig.~\ref{fig:1}(b)], which explains the disentangling effect of our method. For the XXZ chain, CAMPS gives rise to an exact duality mapping the XXZ chain to the quantum Ashkin-Teller spin chain, which is equivalent to two Ising chains coupled together. In particular, for the XX chain, we find that CAMPS transforms the system into two decoupled critical quantum Ising chains [see Fig.~\ref{fig:1}(b)], which locate, respectively, at the left and right half of the chain, significantly reducing the entanglement in the ground state. For a small finite coupling along the $z$-direction, similar results hold for short systems, although there exists a small but finite entanglement at the center bond. We also carefully study the universal entanglement spectrum, further confirming our analysis. Our results show that CAMPS is not only an efficient method for simulating quantum many-body systems by decreasing entanglement, but it is also a helpful tool to uncover dualities in quantum critical chains.           

\begin{figure}[!tbp]
    \centering
    \includegraphics[width=1.0\linewidth]{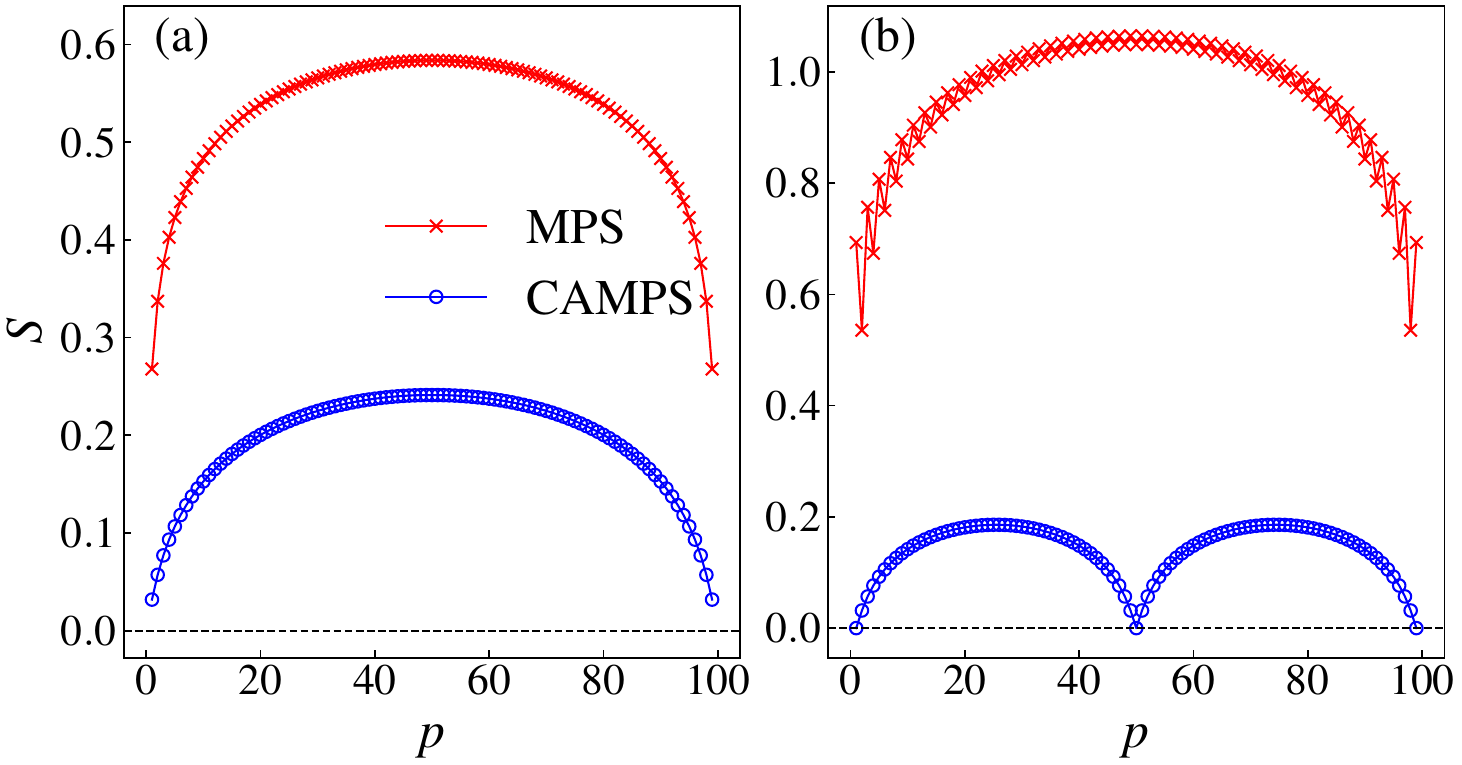}
    \caption{Entanglement entropy at different cuts for the critical Ising model (a) and XXZ chain at $g=0$, i.e., the XX model (b) with open boundary conditions. Notice that the entanglement entropy from CAMPS is actually the NsEE. The size of the chain is $L=100$. The bond dimension is $D=200$ and $60$ for MPS and CAMPS, respectively, which gives converged results. $p$ denotes the position of the cut.}
    \label{fig_entanglement_entropy}
\end{figure}

\section{CAMPS: Disentangling Quantum Many-body Systems with Clifford Circuits}
Here, we give a brief introduction to the CAMPS method; more details can be found in Ref.~\cite{PhysRevLett.133.190402}. The essence of CAMPS is to transfer the so-called stabilizer entropy to Clifford circuits, allowing MPS to handle only the rest of the entanglement entropy, i.e., the so-called non-stabilizerness entanglement entropy \cite{2024arXiv240916895H}. An illustration of the wave-function ansatz of CAMPS can be found in Fig.~\ref{fig:1}(a), in which Clifford circuits are applied to the MPS repeatedly. In CAMPS, the modification of the DMRG algorithm is minor. After obtaining the eigenstate of the effective Hamiltonian, a two-site Clifford circuit is applied before the singular value decomposition (SVD) and the truncation is performed. The criterion for choosing the two-site Clifford circuits is to search for the ones which yield the smallest entanglement entropy~\cite{10.1063/1.4903507,PhysRevB.100.134306,PhysRevA.87.030301}. We notice also that only non-equivalent Clifford circuits (not connected by single qubit gates) need to be considered \cite{2024arXiv241009001F}.  As Clifford circuits preserve the Pauli string structure, the update of the Hamiltonian can be implemented easily. CAMPS provides a seamless integration of Clifford circuits into the DMRG algorithm, optimizing the wave-function ansatz structure and the local tensor simultaneously. In this sense, there is no restriction on the number of layers of Clifford circuits, in contrast to a previous ansatz \cite{Qian_2023}. Previous applications of CAMPS have shown that it is a very effective numerical method for simulating quantum many-body systems \cite{PhysRevLett.133.190402,2024arXiv240703202Q,2024arXiv240701692M,2024arXiv241015709Q}. We notice that, in the special case of the toric code model, CAMPS can find the Clifford circuits which transform its ground state to a direct product state \cite{2024arXiv240916895H}. Moreover, since the application of Clifford circuits does not change the measurement of magic (stabilizer Rényi entropy \cite{PhysRevLett.128.050402}, for example), CAMPS can also serve as an efficient way to calculate measurements of magic \cite{2024arXiv240916895H}. 

As a by-product of CAMPS, a measurement of the hardness for simulating a quantum many-body system classically called non-stabilizerness entanglement entropy (NsEE) \cite{2024arXiv240916895H} was proposed. NsEE is essentially the entanglement entropy which cannot be removed by the Clifford circuits. It is shown to be a better measurement than the entanglement entropy or magic alone. In CAMPS, the MPS needs only to handle the NsEE since the stabilizer entanglement entropy can be captured by the Clifford circuits.

\begin{figure}[!tbp]
    \centering
    \includegraphics[width=1.0\linewidth]{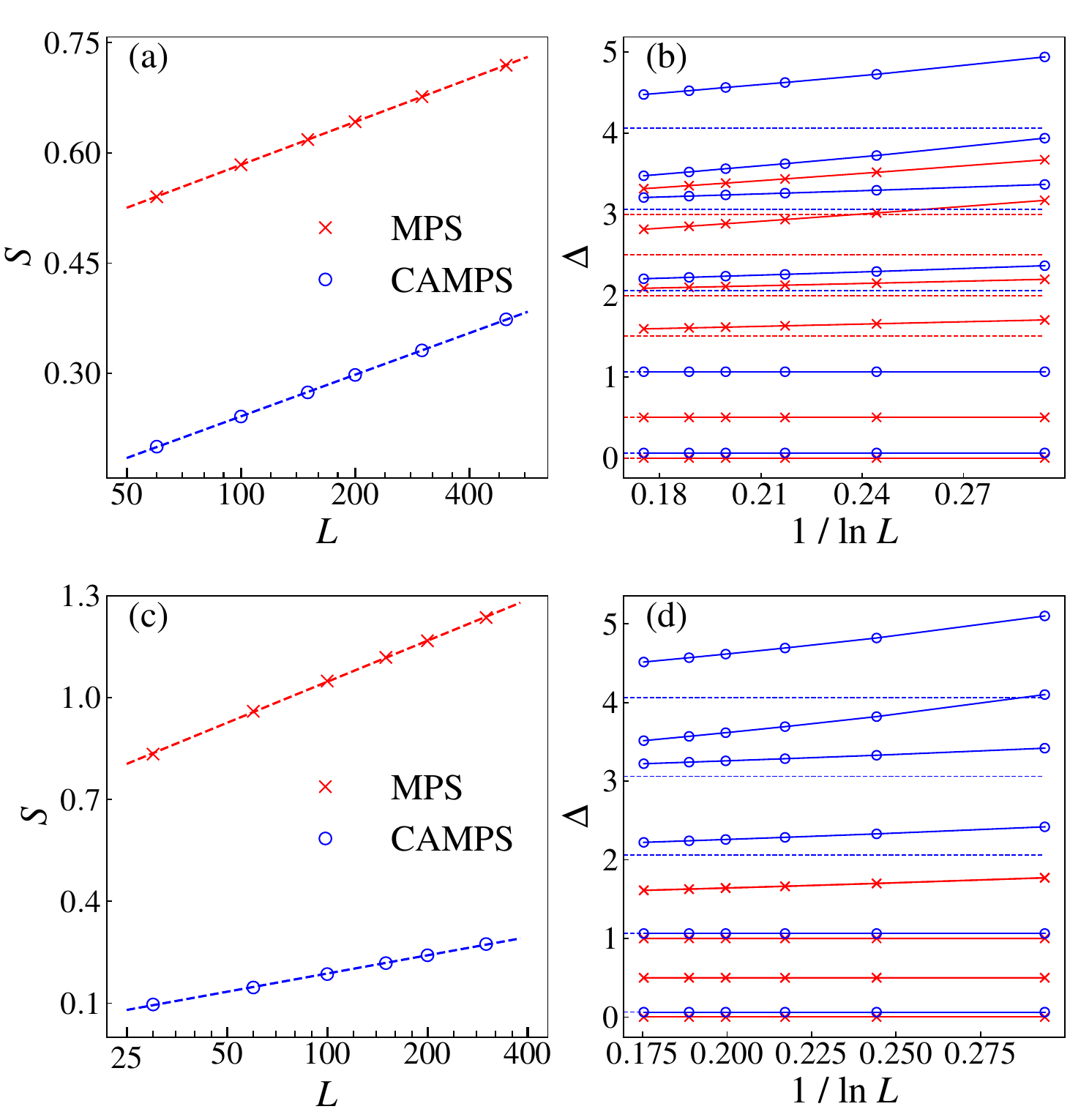}
    \caption{Comparison of the MPS and the CAMPS results for the critical quantum Ising [(a) and (b)] and XX [(c) and (d)] chains. Notice that the entanglement entropy from CAMPS is actually the NsEE. Finite-size scaling of the entanglement entropy is shown in (a) for the critical quantum Ising chain and (c) for the XX chain. Shifted and rescaled entanglement spectrum is shown in (b) for the critical quantum Ising chain and (d) for the XX chain. The cut is at the center bond of the spin chain in all the cases except for the CAMPS result of the XX chain, where the cut is at the $L/4$-th site because the spin chain is transformed into two decoupled Ising chains [see Fig.~\ref{fig_entanglement_entropy}(b)]. In (a) and (c), the $x$-axis is in the logarithmic scale and the central charges are fitted to be $c\sim0.50$ for the CAMPS result. In (b) and (d), the lowest eigenvalue in the entanglement spectrum $\Delta$ is normalized to be $0$ and $1/16$ for the MPS and CAMPS results, respectively. For both MPS and CAMPS, converged results with bond dimension $D$ are plotted.}
    \label{fig_isingandXX}
\end{figure}

\section{Results}

The correlation between two subsystems of a quantum system can be quantified by the entanglement entropy. For a system in state $|\Psi\rangle$ divided into two parts $A$ and $B$, the entanglement entropy with respect to this bipartition is defined as the von Neumann entropy
\begin{equation}
    S_{A/B} = -\mathrm{Tr}(\rho_{A/B} \log \rho_{A/B})
\end{equation}
of the reduced density matrix $\rho_{A/B} = \mathrm{Tr}_{B/A}(\rho)$, where $\rho = \left |\Psi \right \rangle \left \langle \Psi \right |$ is the density matrix of the entire system. It can be easily shown that $S_A = S_B$.

For a critical quantum chain with length $L$ and open boundary conditions described by a CFT, the entanglement entropy scales as~\cite{Cardy2016,PhysRevLett.96.100603} 
\begin{equation}
 S = \frac{c}{6}\ln L + b
\end{equation}
with $c$ the central charge of the underlying CFT and $b$ containing the contribution from the boundaries~\cite{PhysRevLett.96.100603}. In addition to the entanglement entropy, the entanglement spectrum $\Delta$ (defined as $-\ln\lambda$ and $\lambda$ is the eigenvalue of $\rho_A$ or $\rho_B$) can be used to identify the underlying CFT of the critical chain. 

\subsection{Ising CFT: CAMPS changes boundary}
To begin with, we study the quantum Ising chain
\begin{equation} \label{eq_ising}
    H_\text{Ising} = -\sum_{1 \leq j<L} Z_j Z_{j+1} - g \sum_{1 \leq j \leq L } X_j
\end{equation}
as an illustration of our CAMPS approach to critical models. It enjoys an on-site $\Z_2$ global symmetry generated by $\eta := \prod_j X_j$. There is a continuous phase transition at $g=1$ separated by the $\Z_2$ symmetric phase ($g>1$) and spontaneous symmetry-breaking phase ($g<1$). The low-energy physics at the quantum critical point (QCP) is described by the Ising CFT with central charge $c=1/2$. Interestingly, besides the $\Z_2$ symmetry, the QCP has a generalized Kramers-Wannier duality symmetry (\textbf{KW}) acting as 
\begin{equation}
    \begin{aligned}
        & \textbf{KW} \, Z_j Z_{j+1} = X_{j}\, \textbf{KW} \\
        & \textbf{KW} \, X_{j+1} = Z_j Z_{j+1}\, \textbf{KW}.
    \end{aligned}
\end{equation}
One can see that \textbf{KW}, which amounts to gauging the global $\Z_2$ symmetry, maps a symmetric lattice operator to a different one and changes the corresponding phases. \textbf{KW} together with $\eta$ form a generalized fusion category symmetry, which was found to be the key to understanding critical states~\cite{Vanhove_2022,aasen2020topologicaldefectslatticedualities,chen2022,seiberg2024,li2023intrinsically,li2023non,yan2024generalized,lootens2024entanglementdensitymatrixrenormalisation,zhang2024KW}. The Kramers-Wannier self-duality is also important to understand the entanglement properties in critical states. In particular, we will show here that \textbf{KW} is closely related to the reduction of entanglement entropy in CAMPS. 

We first demonstrate the effectiveness of CAMPS for the critical Ising chain ($g=1$) defined in Eq.~\eqref{eq_ising} under open boundary conditions in Fig.~\ref{fig_entanglement_entropy}(a). As expected, the entanglement entropies at different cut locations are all significantly reduced. In particular, the entanglement entropy almost disappears close to the boundary. In Fig.~\ref{fig_isingandXX}(a) and (b), we show the further comparison between MPS and our CAMPS results for the critical Ising chain. We find that in both cases, the entanglement entropy contains a logarithmic leading contribution, which gives a central charge $c \sim 0.50$ from the fitting. This supports that the original model and the model conjugated by the Clifford circuits are both described by the Ising CFT. Hence, the reduction of the entanglement entropy is related to the boundary entropy in a CFT ground state~\cite{PhysRevLett.67.161}. The CAMPS results indicate the entanglement Hamiltonian in the conjugated model is approximately a different boundary CFT (BCFT)~\cite{Laeuchli2013,Cardy2016,PhysRevLett.132.086503,schneider2024selfcongruentpointcriticalmatrix}. This is verified by the universal entanglement spectrum shown in Fig.~\ref{fig_isingandXX}(b). The entanglement Hamiltonian in the MPS contains two conformal towers labeled by $I$ and $\epsilon$ from the spectrum, as it is approximately a BCFT with free boundaries. The CAMPS results, on the other hand, contain only a $\sigma$ conformal tower, indicating that the entanglement Hamiltonian of the conjugated model is a BCFT with mixed boundaries. As a result, we can conclude that the CAMPS reduces the entanglement entropy by changing the free boundary condition to the fixed one in the critical Ising chain. 

Besides the entanglement properties analyzed above, we can actually examine the explicit form of Clifford circuits optimized from CAMPS. We find that it is nothing but \textbf{KW}, the exact duality transformation. The Clifford circuit found in the CAMPS calculation is $\prod_{i=1}^{L-1}\text{CNOT}_{i+1,i}$ (the details can be found in Appendix~\ref{sec:optimized-Clifford-circuits}). We notice that $\prod_{i=1}^{L-1}\text{CNOT}_{i+1,i}$ can be represented as a matrix product operator (MPO) with bond dimension $4$, which can explain why the central charge is not altered after the application of the Clifford circuits. After a local rotation, the Hamiltonian conjugated by the Clifford circuits, or the dual Hamiltonian, to put it another way, is found to be in the same form as the original one, except for the boundary terms
\begin{equation} \label{eq:eq_ising_gauged}
    \begin{aligned}
    \widetilde{H}_\text{Ising} = & - g \sum_{1\leq j< L} Z_j Z_{j+1} -  \sum_{1\leq j< L} X_j - g\,Z_1.
    \end{aligned}
\end{equation}
A $\Z_2$ symmetry-breaking field emerges at one boundary in the conjugated model and a transverse-field term is missing at the other boundary, consistent with previous results~\cite{CARDY1986200,CARDY1989581,CARDY1984514}. It is known that the critical Ising chain under the open boundary conditions can be described by a free BCFT~\cite{CARDY1986200,CARDY1989581,CARDY1984514}. \textbf{KW} maps it to a BCFT with fixed boundary conditions~\cite{ref_ising_bc_discussion}. Accordingly, the entanglement Hamiltonian is approximately a free or mixed BCFT, respectively~\cite{Laeuchli2013,Cardy2016}, for which the boundary entropies differ by a constant term. This explains the usefulness of CAMPS in the study of the critical quantum Ising chain.

\begin{figure}[!tbp]
    \centering
    \includegraphics[width=1.0\linewidth]{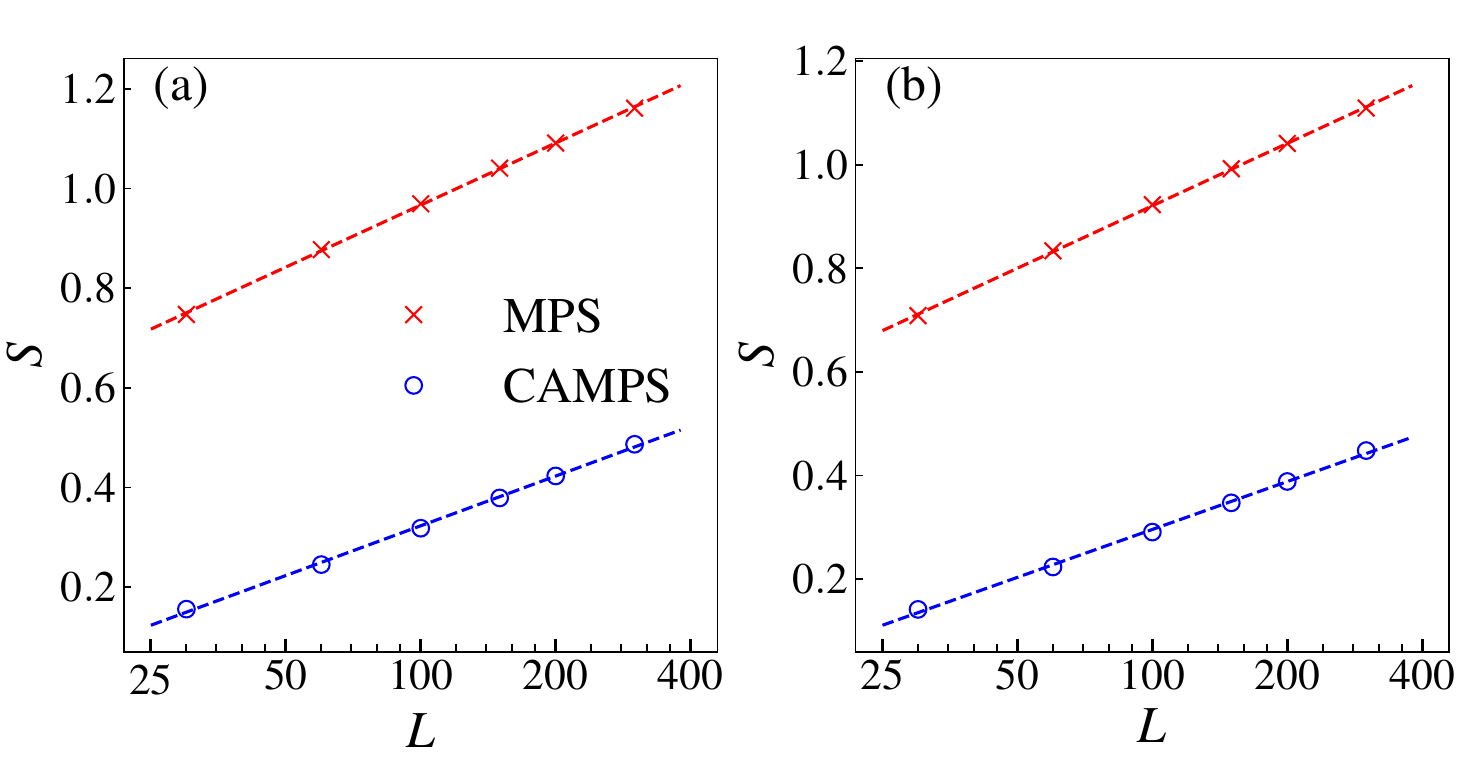}
    \caption{Comparison of the MPS and the CAMPS results for the XXZ chain with (a) $g = 0.5$ and (b) $g=1$. Finite-size scaling of the entanglement entropies are shown. For both cases, only a constant reduction of entanglement entropy is found in CAMPS result, while the slopes remain unchanged. The reduction values can be found in Fig.~\ref{fig_entropyfit} and Table.~\ref{t1} in the Appendix. These results are consistent with the fact that optimized Clifford circuit in CAMPS [see Fig.~\ref{fig_circuitIsingandHei_1}(b) in Appendix~\ref{sec:optimized-Clifford-circuits}] is a MPO with finite bond dimension.
    }
    \label{fig:XXZ}
\end{figure}

\subsection{$c=1$:~Clifford circuit as an intertwiner}
Our example of the critical Ising chain has illustrated the situation where the CAMPS method yields the circuit implementing the Kramers-Wannier {\it self-}duality. More generally, however, the duality transformation (also known as an {\it intertwiner} in this case) maps a critical quantum chain to another model corresponding to a different CFT in the scaling limit. To exemplify this, we turn to the XXZ spin chain with Hamiltonian
\begin{equation}
\label{eq:XXZ-hamiltonian} 
    H_{\text{XXZ}} = \sum_{j=1}^{L-1} \left( X_{j}X_{j+1} + Y_{j}Y_{j+1} + g\,Z_{j}Z_{j+1} \right).
\end{equation}
As special cases, this Hamiltonian reduces to that of the antiferromagnetic (resp. ferromagnetic) Heisenberg XXX chain when the anisotropy $g = 1$ (resp. $-1$) and to that of the XX chain when $g = 0$. In the range $-1 < g \leq 1$, this model is critical and described by the compactified boson CFT with $c=1$ in the scaling limit; the compactification radius depends on $g$~\cite{Giamarchi-Book}.

Using the CAMPS method, we have performed the optimization for the model defined in Eq.~\eqref{eq:XXZ-hamiltonian} under open boundary conditions. The resulting Hamiltonian reads
\begin{align}
\label{eq:AT-hamiltonian}
    \widetilde{H}_{\text{XXZ}} = &\sum_{j=2}^{L-3} X_{j}X_{j+2} + \sum_{j=2}^{L-1} Y_{j} \nonumber \\
    &- g \sum_{l=1}^{\frac{L}{2}-2} X_{2l}X_{2l+1}X_{2l+2}X_{2l+3} \nonumber \\
    &- g \sum_{l=1}^{\frac{L}{2}-1} Y_{2l}Y_{2l+1} \nonumber \\
    &+ X_{1}X_{2} + X_{3} + X_{L-2} + X_{L-1}X_{L} \nonumber \\
    &- g \left( X_{1}X_{2}X_{3} + X_{L-2}X_{L-1}X_{L} \right).
\end{align}
The optimized Clifford circuits can be found in the Appendix. Remarkably, it is precisely the Hamiltonian of the quantum Ashkin-Teller chain up to the boundary terms in the last two lines of Eq.~\eqref{eq:AT-hamiltonian}. It is known that the Ashkin-Teller model can be obtained from the XXZ model via two consecutive Kramers-Wannier duality transformations, one on the entire chain and the other on the even (or odd) sublattice~\cite{PhysRevB.24.5229,PhysRevB.46.3486}; the corresponding CFT resides on the orbifold branch in the $c=1$ theory space~\cite{GINSPARG1988153}. Thus, applied to the XXZ chain, the CAMPS method leads to a variational realization of the intertwiner for these transformations, or, on the level of CFTs, that for the $\mathbb{Z}_2$ orbifolding.

The comparison of the results from the CAMPS method with those from the ordinary MPS simulation is shown in Fig.~\ref{fig_entanglement_entropy}(b). Again, a significant reduction of the entanglement entropy is observed. A particularly interesting case is that of the XX chain, i.e., $g = 0$. Except for the boundary terms, the conjugated Hamiltonian is now that of two completely decoupled critical Ising chains. More interestingly, the variational optimization further rearranges the sites with swap gates to move the even (resp. odd) sites to the left (resp. right) half of the chain (except the first and last sites), thus making the two Ising chains spatially separated, see Fig.~\ref{fig:1}(b).  We have thus recovered, through the variational approach, the famous duality between the CFT of a free boson with the compactification radius equal to $1$ (or $2$, due to the T-duality) and two copies of the Ising CFT. Indeed, the fitting in Fig.~\ref{fig_isingandXX}(c) for CAMPS results yields the central charge $c \sim 0.50$, in agreement with that of (one single copy of) the Ising CFT.

In Fig.~\ref{fig:XXZ}, we show the comparison of the entanglement entropy between MPS and CAMPS for $g = 0.5$ (a) and $g = 1$ (b). For both cases, only a constant reduction of entanglement entropy is found in the CAMPS result, while the slopes remain unchanged. The values of the reduction can be found in Fig.~\ref{fig_entropyfit} and Table.~\ref{t1} in Appendix~\ref{sec:reduced-entanglement-entropy}. These results are consistent with the fact that the optimized Clifford circuit in CAMPS [see Fig.~\ref{fig_circuitIsingandHei_1}(b) in Appendix~\ref{sec:optimized-Clifford-circuits}] is an MPO with finite bond dimension.

\section{Conclusion and Perspectives}
We have studied the disentangling power of Clifford circuits on critical spin chains employing the CAMPS method. For both the critical quantum Ising chain and the XXZ chain, Clifford circuits can reduce the entanglement significantly. For the critical quantum Ising chain, CAMPS finds the Kramers-Wannier self-duality, which changes the boundary condition of the spin chain. For the XXZ chain, CAMPS produces its duality transformation into the Ashkin-Teller chain. In particular, for the special case of the XX chain, the Clifford circuits resulting from CAMPS transform the model into two decoupled critical quantum Ising chains located at the left and right halves of the lattice. The disentangling power of CAMPS can thus be understood as a variational approach to finding duality transformations that make the critical models less entangled.

It is interesting that the duality in both the transverse Ising and XXZ models can be represented by Clifford circuits. There might exist other dualities whose representations are beyond Clifford circuits. However, given that Clifford circuits preserve Pauli string, duality which connects two local models with the same number of local terms is highly
likely to be able to be represented by Clifford circuits. The framework proposed in this work offers a valuable tool to disentangle critical chains and unveil the hidden dualities. Besides known critical chains with exact duality properties, it is tantalizing to apply this framework to other critical models to discover possible new dualities. In this work, we have been focusing on the one-dimensional critical quantum chains; it would also be interesting to investigate the consequential Clifford circuits and the conjugated Hamiltonians for two-dimensional systems.

{\it Note added.}
Upon completing our work, we became aware of Ref.~\cite{frau2024stabilizerdisentanglingconformalfield}, where the application of CAMPS to quantum critical chains was also studied. Our analyses are, however, quite different, as our focus has been on the duality transformations of the underlying conformal field theories.

{\it Acknowledgements:}
We thank Linhao Li and Atsushi Ueda for helpful discussions. M. P. Qin acknowledges the support from the National Key Research and Development Program of MOST of China (Grant No. 2022YFA1405400), the National Natural Science Foundation of China (Grant No. 12274290), the Innovation Program for Quantum Science and Technology (Grant No. 2021ZD0301900), and the sponsorship from Yangyang Development Fund. R. Z. Huang is supported by a postdoctoral fellowship from the Special Research Fund (BOF) of Ghent University. H.-C. Zhang acknowledges support from the Spanish MINECO Grant No. PID2021-127726NB-I00, the CSIC Research Platform on Quantum Technologies PTI-001, and the QUANTUM ENIA project Quantum Spain through the RTRP-Next Generation program under the framework of the Digital Spain 2026 Agenda. T. Xiang is supported by the National Key Research and Development Program of MOST of China (Grant No. 2021ZD0301800) and the National Natural Science Foundation of China (Grant No. 12488201).

\appendix

\section{Energy comparison}
\begin{figure}[!tbp] 
    \centering
    \includegraphics[width=1.0\linewidth]{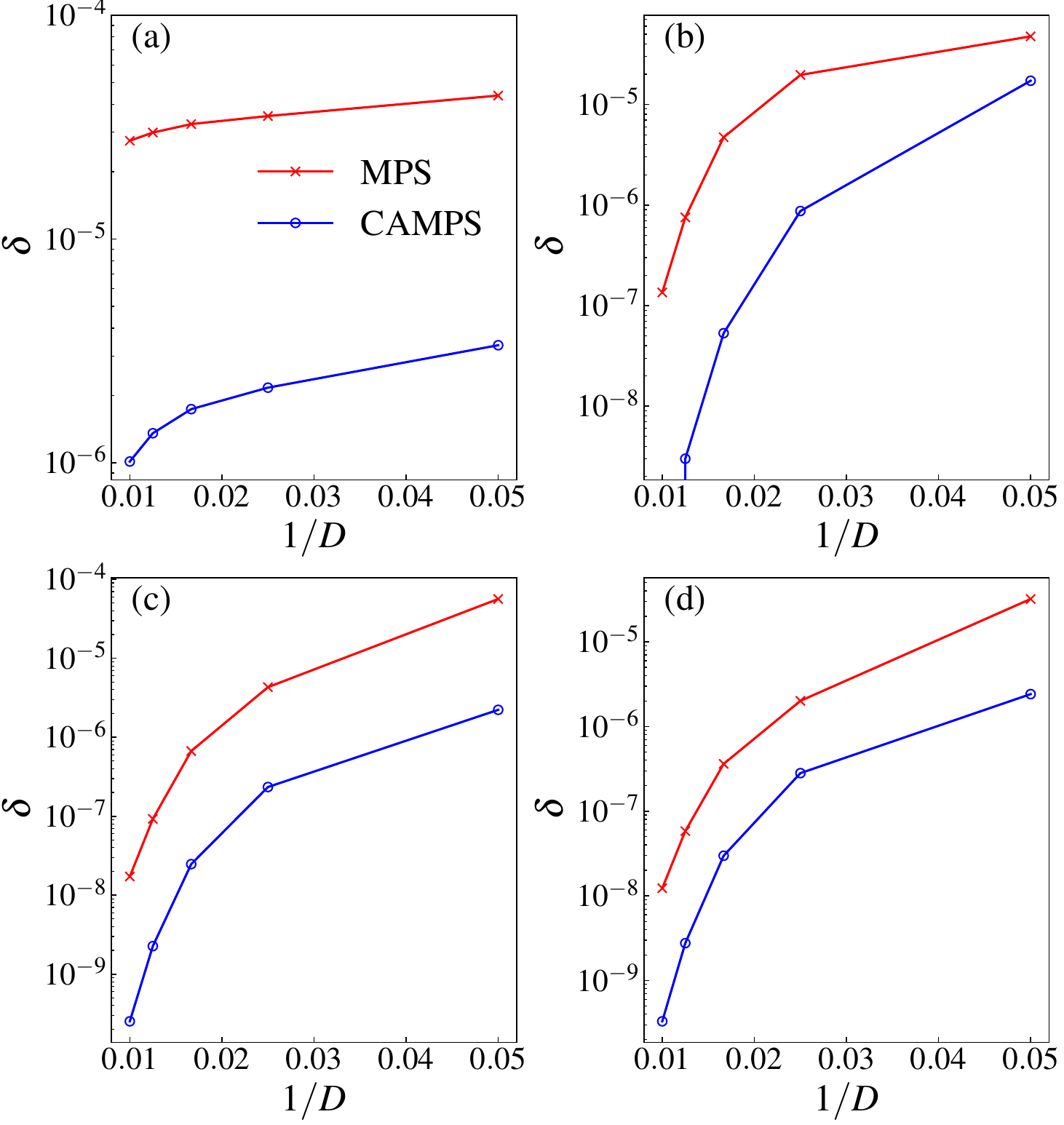}
    \caption{Comparison of the relative error of calculated ground state energy ($\delta$) between MPS and CAMPS for the (a) critical Ising (b) XX, (c) XXZ with $g=0.5$, and (d) Heisenberg model. The system size is $L = 200$. The MPS value of $D=1000$ is used as reference.}
\label{fig_energycomparison}
\end{figure}

In Fig.~\ref{fig_energycomparison}, we show the comparison of the ground state energy obtained by MPS (DMRG) and CAMPS, for critical Ising (a), XX (b), XXZ with $g=0.5$ (c), and Heisenberg (d) models with $L=200$. One finds a significant improvement in accuracy of CAMPS over MPS for the same bond dimension.

\section{The reduced entanglement entropy with CAMPS in the thermodynamic limit}
\label{sec:reduced-entanglement-entropy}

\begin{figure}[!tbp] 
    \centering
    \includegraphics[width=1.0\linewidth]{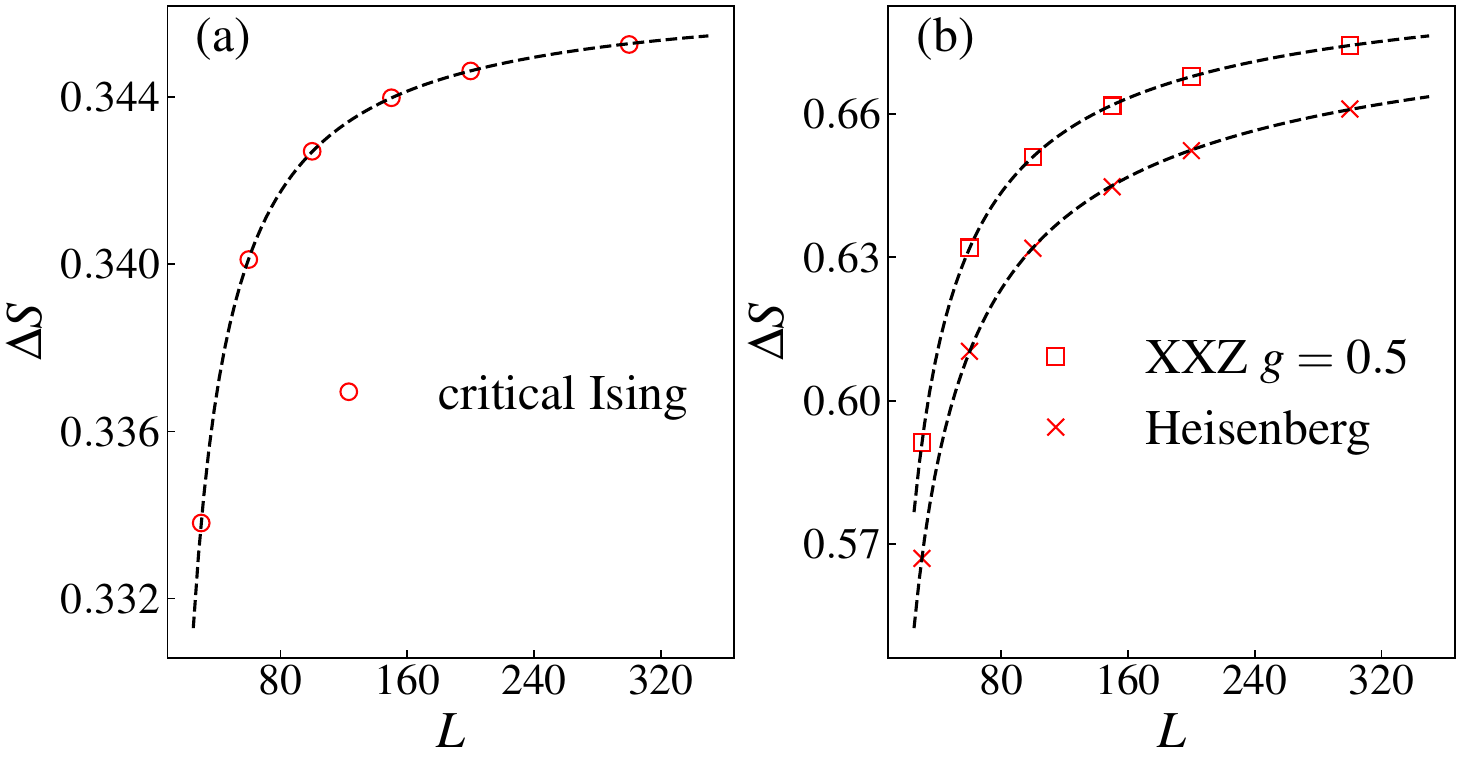}
    \caption{The reduction of entanglement entropy in CAMPS ($\Delta S$) for (a) the critical Ising (b) the XXZ with $g=0.5$, and the Heisenberg models. Fits using $\Delta S = \alpha \times L^{-\beta} + \gamma$ are also shown.}
\label{fig_entropyfit}
\end{figure}

For both the critical Ising and XXZ chains (except for the case of the XX chain), CAMPS can only reduce the contribution of entanglement entropy from boundaries and the central charge remains unchanged. In Fig.~\ref{fig_entropyfit}, we plot the reduction of entanglement entropy with CAMPS (denoted as $\Delta S$) and fit it using $\Delta S = \alpha \times L^{-\beta} + \gamma$, in which $\beta$ is a positive number and $\gamma$ is the reduction of entanglement entropy with CAMPS in the limit of $L\rightarrow \infty$. The fitted results of $\gamma$ for different models are listed in Table.~\ref{t1}.

\begin{table}[h!] 
    \caption{Reduction of entanglement entropy with CAMPS in the thermodynamic limit.}
    \begin{tabular}{p{2cm}|>{\centering\arraybackslash}p{3cm}}
    \hline
    Model & $\gamma$ \\ 
    \hline
    Critical Ising   & 0.3466(1)\\ 
    XXZ $g=0.5$   &  0.6931(14)\\ 
    Heisenberg   &  0.6906(30)\\ 
    \hline
    \end{tabular}
\label{t1}
\end{table}

\begin{figure}[b]
    \centering
    \includegraphics[width=1.0\linewidth]{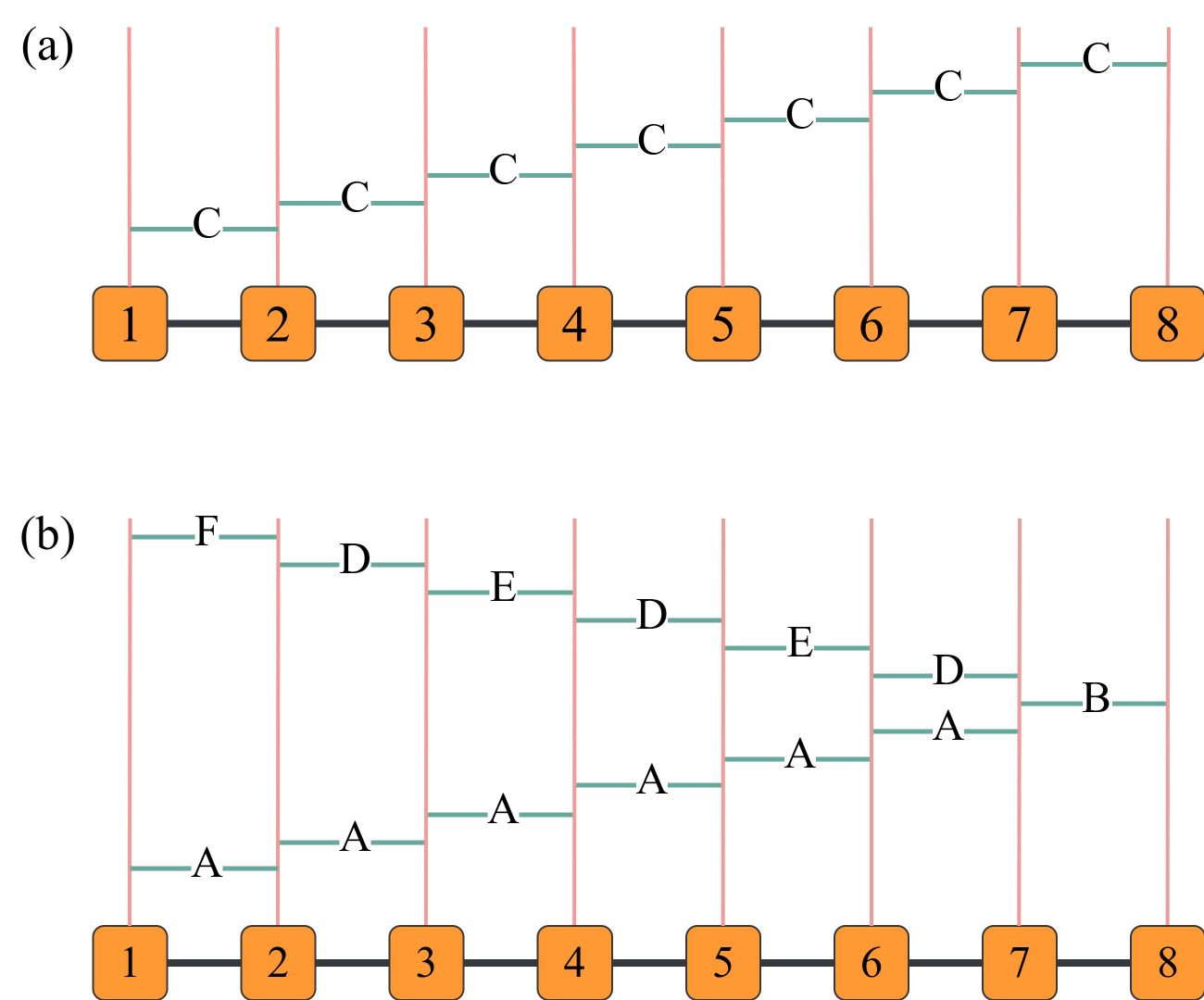}
    \caption{Illustration of the optimized Clifford circuits for the critical transverse Ising (a) and XXZ (b) chains. Here, systems of $8$ sites are used as examples. Results for larger system size are similar. As mentioned in the main text, the optimized Clifford circuits are $\prod_{i=1}^{L-1}\text{CNOT}_{i+1,i}$ for the critical transverse Ising chain. The optimized Clifford circuits for XXZ are more complicated, as shown in (b), where the effect of the two-qubit circuits is listed in Table.~\ref{def_gate}.}
    \label{fig_circuitIsingandHei_1}
\end{figure}

\section{The optimized Clifford circuits using CAMPS}
\label{sec:optimized-Clifford-circuits}

\begin{figure}[b]
    \centering
    \includegraphics[width=1.0\linewidth]{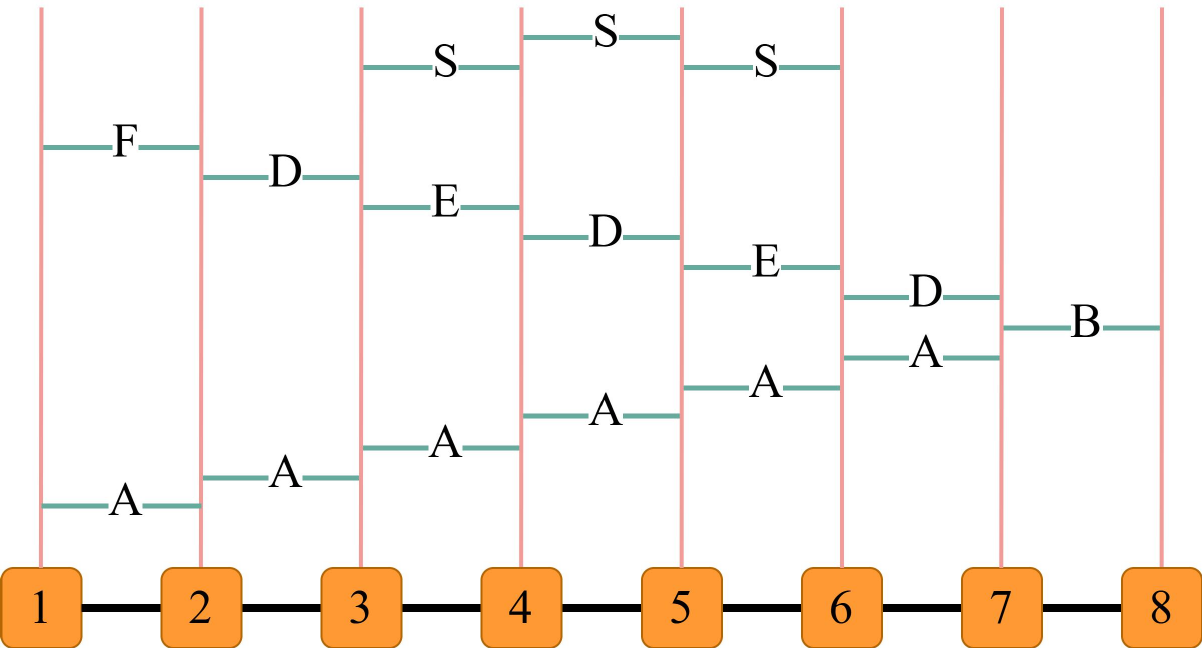}
    \caption{Illustration of the optimized Clifford circuits for the XX chain. The first two layers are exactly the same as the XXZ chain in Fig.~\ref{fig_circuitIsingandHei_1}(b). CAMPS also finds layers of swap gates in the shape of a pyramid which arrange even (odd) sites into the left (right) of the chain (except for the first and the last sites), spatially separating the two decoupled critical transverse Ising chains.}
    \label{fig_circuitXX_1}
\end{figure}

As mentioned in the main text, for the critical transverse Ising chain, the optimized Clifford circuits from CAMPS are $\prod_{i=1}^{L-1}\text{CNOT}_{i+1,i}$ [illustrated in Fig.~\ref{fig_circuitIsingandHei_1}(a)]. The optimized Clifford circuits for the XXZ model are more complicated [illustrated in Fig.~\ref{fig_circuitIsingandHei_1}(b)]. The effect of each two-qubit Clifford circuits in Fig.~\ref{fig_circuitIsingandHei_1}(b) is listed in Table.~\ref{def_gate}. For the special case of the XX chain, where the corresponding Ashkin-Teller chain becomes two decoupled critical transverse Ising chains, CAMPS finds layers of swap gates in the shape of a pyramid which arrange even (odd) sites into the left (right) of the chain (except for the first and last sites; see Fig.~\ref{fig_circuitXX_1}). In the XX case, the number of layers of the optimized Clifford circuits scales linearly with $L$, which is responsible for the change of central charge from $c = 1$ in the XX case to $c = 1/2$ in the critical transverse Ising case.

\begin{table}[h!]
    \caption{Gates used in circuits in Fig.~\ref{fig_circuitIsingandHei_1}.}
    \begin{tabular}{c|c|c|c|c|c|c|c}
    \hline
    Label & A & B & C & D & E & F & S \\ 
    \hline
    XI   & +ZZ   & +XZ & +XI  & -IX   & +XI   & +ZI & +IX   \\ 
    IX   & +IZ   & +ZX & +XX  & +YI   & -XZ   & -ZZ & +XI   \\ 
    ZI   & +YZ   & +XY & +ZZ  & -IZ   & +ZX   & +XX & +IZ  \\ 
    IZ   & +XX   & +YX & +IZ  & +XI   & +IX   & +IX & +ZI  \\ 
    \hline
    \end{tabular}
\label{def_gate}
\end{table}

\clearpage

\bibliography{camps-critical-v2.bib}

\begin{thebibliography}{52}
\bibitem [{ref()}]{ref_ising_bc_discussion}%
  \BibitemOpen
  \href@noop {} {\bibinfo  {journal} {More precisely, the conjugated Ising
  model has one fixed boundary. The other one spontaneously breaks the global
  symmetry, hence the up and down fixed boundary conditions both contribute to
  the partition function. In the variational simulation of a quantum chain, a
  fixed boundary is pinned down}\ }\BibitemShut {NoStop}%
\end{thebibliography}

\end{document}